\documentclass[a4paper,11pt]{article}
\usepackage{pos}
\usepackage{lineno}
\title{Precision luminosity measurement with proton-proton collisions at the CMS experiment in Run 2}
\ShortTitle{Precision luminosity measurement with pp collisions at the CMS experiment in Run 2}

\author*[1]{Angela Giraldi}

\affiliation{Deutsches Elektronen-Synchrotron (DESY),\\
  Notkestrasse 85, Hamburg, Germany}
\note{For the CMS Collaboration.}

\emailAdd{angela.giraldi@cern.ch}

\newcommand{\lumi}{\mathcal{L}}
\newcommand{\sigmavis}{\ensuremath{\sigma_{\mathrm{vis}}}}

\abstract{

A precision luminosity measurement is essential for LHC cross-section measurements to determine fundamental parameters of the standard model and constrain or discover beyond-the-standard-model phenomena.
The luminosity of the CMS detector has been measured at the LHC Interaction Point~5 using proton-proton collisions at $\sqrt{s}=13$ TeV during the Run 2 data-taking period \mbox{(2015-2018)}. The absolute luminosity scale is obtained using beam-separation scans and the Van der Meer (VdM) method, and several systematic uncertainty sources are investigated, from the knowledge of the scale of beam separation provided by LHC magnets to the nonfactorizability of the spatial components of proton bunch density distributions in the transverse direction. When the VdM calibration is applied to the entire data-taking period, the detector linearity and stability measurements contribute significantly to the total uncertainty in the integrated luminosity. In 2016-2018, the reported integrated luminosity was among the most precise measurements at bunched-beam hadron colliders. 

}

\FullConference{%
  41st International Conference on High Energy physics - ICHEP2022\\
  6-13 July, 2022\\
  Bologna, Italy
}

\begin{document}
\renewcommand{\logo}{\relax}
\maketitle

\section{Introduction}
The luminosity $\lumi$ is a fundamental property of any collider experiment as it is a measure of the collision rate. It is the proportionality factor that relates the event rate $dN/dt$ of a process to its visible cross-section $\sigma$:
\begin{equation}
    \frac{dN}{dt} = \lumi \cdot \sigma.
    \label{eq:rate}
\end{equation}
%The rate as well as the luminosity are continuously measured online, providing information to the LHC control rooms to optimise data-taking run conditions.\\
Precision cross-section measurements require an accurate understanding of the luminosity integrated over time, mainly because the uncertainty in the integrated luminosity is an important uncertainty in most precise measurements.

The CMS experiment~\cite{bib:CMS} employs a two-step strategy to measure the integrated luminosity of its collision data sets in Run 2 (2015-2018).

First various luminosity detectors are calibrated in a dedicated special run. The calibration constant called \textit{visible cross section} \sigmavis\ relates the measured rate of the luminometer to the luminosity at the time of the measurement. Second, the rate measurement is integrated over the data collection period and normalized with \sigmavis\ to yield the integrated luminosity. 

The methodology and dominant sources of systematic uncertainty are discussed in these proceedings, based on the latest luminosity measurements from proton-proton (pp) collision datasets collected at $\sqrt{s}=13$ TeV~\cite{bib:CMSlumi1516, bib:CMS2017, bib:CMS2018}.
\section{Luminosity detectors}

During the LHC Run 2 period, the CMS experiment employed several sub-detectors to monitor and measure the luminosity (\textit{luminometers}). There are two dedicated luminosity measurement systems: the Pixel Luminosity Telescope (PLT)~\cite{bib:plt}, consisting of eight three-layer telescopes located \mbox{1.8 m} from the interaction point in both directions, and the Fast Beam Conditions Monitor (BCM1F), made of silicon and diamond sensors with a time resolution of 6.25 ns mounted on the same carriages as PLT. Like the luminosity system installed as part of the Hadron Forward Calorimeter (HF), they use a separate data acquisition (DAQ) system that runs independently of the primary CMS readout. HF uses two algorithms, one based on the transverse energy sum (HFET) and the other on the fraction of occupied towers (HFOC). The results presented here are based primarily on offline measurements made with the forward hadron calorimeter and the pixel cluster counting (PCC) method, using silicon pixels to determine the offline luminosity. 
In addition, other methods are used to perform luminosity measurements using the main CMS DAQ system.  First, the inner tracker information is employed in the vertex counting (VTX) method, with strip tracker information used for cross checks in low-pileup conditions. Then, the rate of the muon track stubs in the muon barrel track finder is measured from the barrel drift tubes (DT), and similarly, the rate of photons in ionization chambers from the CERN radiation and environmental monitoring system (RAMSES) is measured. DT and RAMSES are not sensitive to bunch-by-bunch measurements, and their rates are cross-calibrated in special scans, resulting in stable luminosity measurements during data collection.

\begin{figure}[!b]
    \centering
    \includegraphics[width=0.43\textwidth]{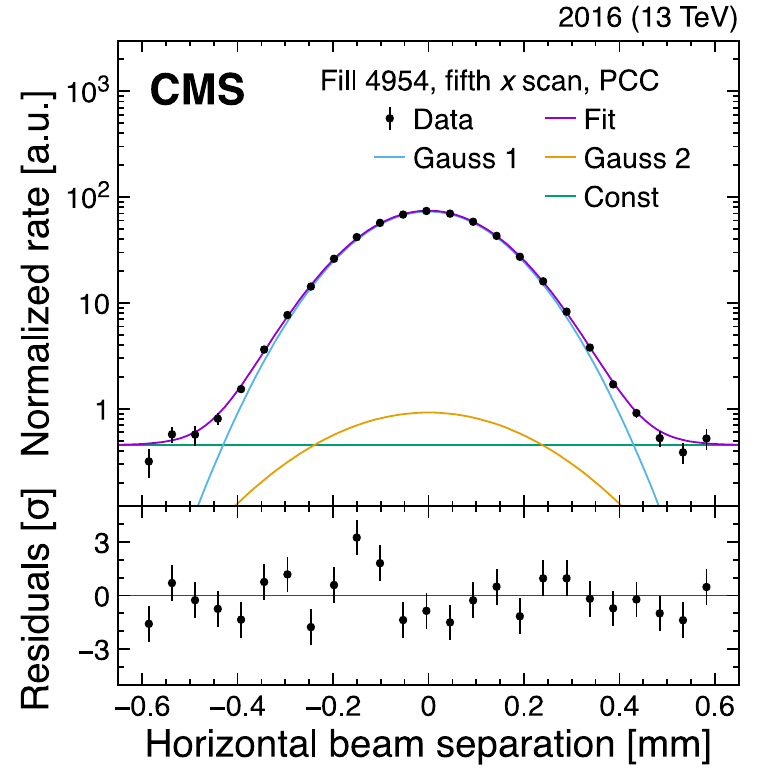}
    \includegraphics[width=0.56\textwidth]{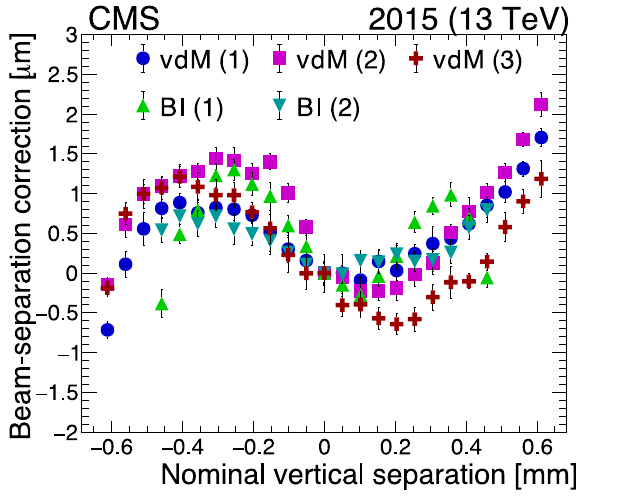}
    \caption{Left: Example of a VdM scan illustrating the normalized rate as a function of beam separation in the \textit{x} direction and the fitted curves. The difference between the measured and fitted values is shown in the lower panel, divided by the statistical uncertainty~\cite{bib:CMSlumi1516}. Right: The beam-separation residuals in \textit{y}~\cite{bib:CMSlumi1516}.}
    \label{fig:fitvdm}
\end{figure}

Each luminometer reads out a rate $\mathcal{R}(t)$ of specific observables (hits, tracks, clusters, etc.), proportional to the instantaneous luminosity $\lumi (t)$ via Eq. \ref{eq:rate} through \sigmavis.
The luminosity detectors are calibrated once per data-taking period in a special LHC fill by calculating \sigmavis\ as:
\begin{equation}
    \sigma_{\mathrm{vis}} = \frac{2 \pi \Sigma_x \Sigma_y}{N_1 N_2 \nu_{\mathrm{LHC}}} \cdot \mathcal{R}_0 \ ,
    \label{eq:sigmavis}
\end{equation}
assuming that the bunch proton density function is factorizable into independent $x$ and $y$ terms; with $\Sigma_x$ and $\Sigma_y$ being the beam overlap width and height of the transverse luminous region where the collisions occur, $N_1$ and $N_2$ the numbers of protons in the two colliding bunches, $\nu_{\mathrm{LHC}}$ the revolution frequency, and $\mathcal{R}_0$ the measured rate for head-on beam collisions. The calibration constant \sigmavis\ is determined from Van der Meer (VdM) beam separation scans ~\cite{bib:vdm}, where the two proton beams are separated in the transverse plane and moved across each other in steps. $\Sigma_x$ ($\Sigma_y$) is obtained as the width of the fitted double-Gaussian function to the measured rate of the VdM scan data as a function of beam separation in the horizontal (vertical) plane. An exemplary fit result is shown in Fig.~\ref{fig:fitvdm} (left).

\section{Systematic uncertainties}
The value of \sigmavis\ derived from the VdM scan analysis is affected by several systematic effects known as \textit{calibration} uncertainties. In contrast, \textit{integration} uncertainties result from detector operations throughout the year and from transferring the calibration of the luminometers measured in the special VdM fill to the conditions in normal physics data taking fills.

\subsection{Calibration uncertainties}
The beam separation in the fit is derived from the current settings of the LHC steering magnets (nominal separation) and is susceptible to several effects. Via length scale scans (LSC), the absolute beam separation is calibrated by comparing nominal beam positions to the measured one of reconstructed interaction vertices with the CMS tracker. Using LHC beam position monitoring (BPM) systems, a time-dependent movement of the proton beams away from their standard orbit (orbit drift) is observed and accounted for both in the VdM overlap fit and in the LSC procedure.

The electromagnetic interactions between the beams manifest themselves through two beam-beam effects. First, due to electric repulsion, the transverse distance between the beams increases and the magnitude of this deflection relies on the nominal beam separation ~\cite{bib:beambeam, bib:beambeamdefelction}. Then, due to focusing and defocusing (dynamic $\beta^*$ effect), the transverse shape of the proton bunches varies, resulting in a separation-dependent rise in the collision rate. 

The proton numbers of the beams are accounted for in the VdM fit by normalizing the measured rate with the beam currents as measured with LHC devices and corrected for contributions from spurious charges in nominally empty bunch slots (\textit{ghost charges}) and non-colliding buckets of the colliding bunches (\textit{satellite charges}).

Several VdM scan pairs are conducted during a calibration. The bunch-to-bunch and scan-to-scan consistency, also using different luminometers, is examined to evaluate the systematic uncertainty of cross-detector calibration.
After applying all the corrections mentioned above, a residual difference is consistently observed, see Fig.~\ref{fig:fitvdm} (right); corrections are taken into account.

The assumption that the bunch proton density function is factorizable into independent horizontal and vertical components, and hence \mbox{Eq. (\ref{eq:sigmavis})} holds, is tested to estimate the bias in the measurement of the beam overlap integral. Different techniques are used to measure this effect, based on beam imaging and offset scans ~\cite{bib:CMSlumi1516, bib:Beamimaging,bib:CMSnonfactorization}.

\subsection{Integration uncertainties}
After determining the calibrated \sigmavis, a measurement of the integrated luminosity can be assessed in the standard physics fills for all data-taking periods.

Most detector measurements involve false signals from \textit{out-of-time pileup} contributions caused by electronic spillover or activation of the surrounding detector material after collisions occurred. It does not affect the calibration but results in a nonlinear detector response under normal data-taking conditions.   The observed rate from nominally empty bunches is used to calculate these afterglow corrections. 

\begin{figure}[!t]
    \centering
    \includegraphics[width=0.7\textwidth]{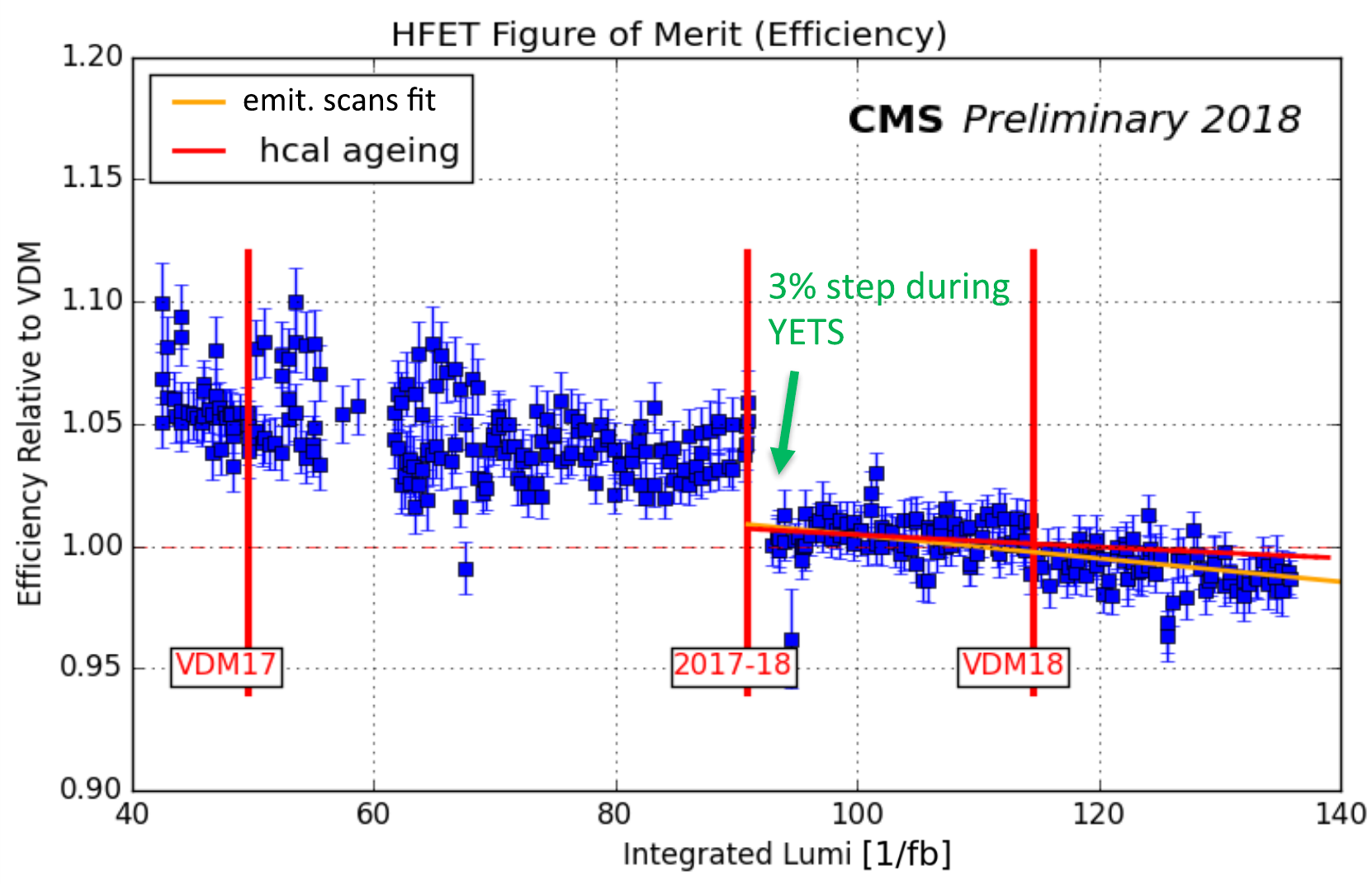}
    \caption{Relative efficiency of the HFET detector response as determined from emittance scans as a function of integrated luminosity for all LHC fills in 2017 and 2018 pp data. The slope indicates the instability~\cite{bib:emittance}.}
    \label{fig:emittance}
\end{figure}

Uncertainties in the integration procedure mainly arise from extrapolating the calibration from the special to regular data-taking conditions (\textit{linearity}), as well as from changes in the  detector over time (\textit{stability}). Different operational difficulties and radiation damage can impact individual luminometers; therefore, their stability and linearity must be monitored and, if feasible, independently corrected. This is accomplished by performing short VdM-like beam separation scans in typical physics conditions at the beginning and end of fills, so-called emittance scans~\cite{bib:emittance}. An example of changing detector conditions over time is the ageing of PMTs and fibers in the forward hadron calorimeter, which causes instabilities in the HFET and HFOC response. The efficiency of the HFET response as evaluated by emittance scans is shown in Fig.~\ref{fig:emittance} for 2017 and 2018, with a slope indicating a significant decrease in efficiency with time.

After applying all corrections, the overall consistency of the individual luminometers values to each other is evaluated, and the residual \textit{cross-detector stability and linearity} uncertainty is assigned. In addition, the uncertainty associated with the deadtime of the CMS data acquisition system is also included.

\begin{table}
  \centering
  \begin{tabular}{lccccc}

    \multicolumn{2}{c}{Source}                                       & 2015  & 2016  & 2017  & 2018  \\
    \hline
    \hline
    \multicolumn{6}{c}{Calibration uncertainty} \\
    Ghost and satellite charge & [\%]      & 0.1 & 0.1  & 0.1  &  0.1 \\
    Beam current normalization  & [\%]     & 0.2 & 0.2 &  0.3 & 0.2  \\
    Orbit drift        & [\%]              & 0.2 & 0.2 & 0.2 & 0.2   \\
    Residual differences    & [\%]         & 0.8 & 0.5 & 0.2 & 0.1  \\
    Beam-beam effects     & [\%]           & 0.5 & 0.5 & 0.6 & 0.2  \\
    Length scale calibration  & [\%]       & 0.2 & 0.3 & 0.3 & 0.2   \\
    Transverse factorizability  & [\%]     & 0.5 & 0.5 & 0.8 & 2.0  \\
    \multicolumn{6}{c}{Integration uncertainty} \\

    Out-of-time pileup Type 1 corrections   & [\%]            & 0.3 & 0.3 & 0.2 & 0.1\\
    Out-of-time pileup Type 2 corrections    & [\%]           & 0.1 & 0.3 & 0.3 & 0.4\\
    Cross-detector stability    & [\%]     & 0.6 & 0.5 & 0.5 & 0.6   \\
    Linearity              & [\%]          & 0.5 & 0.3 & 1.5 & 1.5   \\
    CMS deadtime         & [\%]            & 0.5 & $<$0.1 & 0.5 &  $<$0.1 \\
    \hline 
    Total uncertainty              & [\%]               & 1.6 & 1.2 & 2.3 & 2.5  \\

    Integrated luminosity & [fb$^{-1}$] & 2.27 & 36.3 & 41.5 & 59.8 \\

  \end{tabular}

  \caption{
    Summary of main contributions (in \%) to the systematic uncertainty in \sigmavis\ at $\sqrt{s}$=13 TeV~\cite{bib:CMSlumi1516,bib:CMS2017,bib:CMS2018}. }
  \label{TAB:SystematicError}
\end{table}

\section{Summary}
The integrated luminosity measurement of the CMS experiment is calibrated with the Van der Meer (VdM) approach, considering several systematic uncertainties sources, and corrected for nonlinearity and instability of the detector response. Table~\ref{TAB:SystematicError} summarizes the sources of systematic uncertainty in the luminosity calibration for the pp collision data sets at $\sqrt{s}$ = 13 TeV due to the VdM scan procedure (\textit{calibration}) and the detector operations (\textit{integration}).

\newpage

\bibliographystyle{JHEP}
\bibliography{main}

\end{document}